\documentclass[11pt]{article}
\usepackage{amsmath,amssymb}
\usepackage{hyperref}

\newcommand{\be}{\begin{equation}}
\newcommand{\ee}{\end{equation}}
\newcommand{\bestar}{\begin{equation*}}
\newcommand{\eestar}{\end{equation*}}

\newcommand{\bi}{\begin{itemize}}
\newcommand{\ei}{\end{itemize}}
\newcommand{\bea}{\begin{eqnarray}}
\newcommand{\eea}{\end{eqnarray}}

\newcommand{\hbo}{\hbox to 1 true cm {\hfill } }

\newcommand{\ud}{\mathrm{d}}

\newcommand{\e}{\mathrm{e}}		

\title{Unparticle actions and gauge invariance}

\author{Anton Ilderton\\ School of Mathematics, Trinity College, Dublin 2, Ireland\\
\href{mailto:antoni@maths.tcd.ie}{\texttt{antoni@maths.tcd.ie}}}
\date{}
\begin{document}
\maketitle
\begin{abstract}
\noindent
We show that the requirement of gauge invariance is not enough to fix the form of interactions between unparticles and gauge fields, thus revealing a wide new class of gauged unparticle actions. Our approach also allows us to construct operators which create gauge invariant coloured unparticles. We discuss both their perturbative and non--perturbative properties.
\end{abstract}

\section{Introduction}
The idea that the standard model could contain an as yet unseen, exactly scale invariant sector has received a great deal of attention following the papers \cite{Georgi1, Georgi2}. The new degrees of freedom involved cannot be particles, as scale invariance forbids any definite non--zero mass. This `unparticle stuff' would be quite unlike anything in the standard model we know, but evidence for its existence could be inferred from missing energy signatures corresponding to a \emph{non--integer} number of particles. For details of unparticle phenomenology see the recent reviews \cite{Cheung:2008xu,Rajaraman:2008qt} and references therein.

There has been some debate about how to give standard model quantum numbers, in particular colour charge, to unparticles \cite{Cacciapaglia:2007jq, Licht:2008ic}. The introduction of gauge invariant interaction terms to the nonlocal unparticle action has followed ideas familiar from particle physics; a Wilson line is threaded between the unparticle fields to create a gauge invariant object. The debate originates in whether one should specify a path for the Wilson line or use a particular functional form motivated by the minimal coupling prescription. The first approach is unambiguous, but leads to much more complicated Feynman rules than the second, which initially appears ill defined and requires a more careful mathematical treatment.

These two approaches will be briefly reviewed in Sec.~\ref{rev-sec}. We will show in Sec.~\ref{build-sec}, though, that they only scrape the surface of what is possible. After satisfying the requirements of gauge invariance exactly, we will see that an infinite number of physical degrees of freedom remain available to the action, making infinitely many gauged unparticle theories possible.  This approach will allow us to write down the most general coupling to U(1) gauge fields explicitly, reveal the true role of the `Mandelstam condition' employed in \cite{Cacciapaglia:2007jq}, and give examples of new unparticle actions and their Feynman rules. In Sec.~\ref{na-sec} we extend our results to SU($N$). We also use our methods to write down operators creating gauge invariant colourless and coloured unparticle states, in perturbation theory, and we point out the non--perturbative obstacles to such constructions. In Sec.~\ref{con-sec} we give our conclusions.
\section{Gauging the unparticle action}\label{rev-sec}
As our methods apply equally to fermionic and bosonic unparticles we restrict our discussion, for clarity, to the latter. For an unparticle field $\phi(x)$ with scaling dimension $d$, so $\phi(x)\to \lambda^{d}\phi(\lambda x)$ under scale transformations, the propagator is \cite{Georgi1}
\be\label{theprop}
	\Delta(p) = \frac{i\, a_{2-d}}{(p^2+i\epsilon)^{2-d}}\;,
\ee
where $a_{2-d}$ is a normalisation. To restrict the effects of scale invariance to high energy, one may also include an infrared cutoff (mass term) in the propagator. The effective, nonlocal, and scale invariant action generating the propagator is
\be\label{U-act}
	S_\mathcal{U} = \int\!\ud^4z\ud^4y\ \phi^\dagger(z)\, i\Delta^{-1}(z-y)\,\phi(y)\;.
\ee
It is easily verified using (\ref{theprop}) that the action is scale invariant. Note that at scaling dimension $d=1$, the action and propagator reduce to those of an ordinary free scalar field with a particle interpretation.

The unparticles may be coupled to gauge fields by introducing a path ordered Wilson line into the action,
\be\label{gauged-act}
	S_\mathcal{U}\to S = \int\!\ud^4z\ud^4y\ \phi^\dagger(z)\, i\Delta^{-1}(z-y)\, \mathbb{P}\,\exp\bigg[ -ie \int\limits_y^z \ud w^\mu A_\mu(w)\bigg] \,\phi(y)\;.
\ee
This action is invariant under the gauge transformations\footnote{So that SU($N$) and U(1) results are more directly comparable, the coupling constants in both theories will be denoted by $e$, and our SU($N$) generators $T^a$ are hermitian.}
\bestar
\phi \to U^{-1}\phi\;,\qquad \phi^\dagger\to \phi^\dagger U\;,\qquad A_\mu \to U^{-1}A_\mu U+\frac{1}{ie}U^{-1}\partial_\mu U\;,
\eestar
as the Wilson line, $W(z,y)$, transforms as
\be\label{cod}
W(z,y)\to U^{-1}(z)\,W(z,y)\,U(y)\;.
\ee
Two types of Wilson line have been employed to date. We discuss them below.
\subsection{The Wilson line with Mandelstam condition}
The approach developed in \cite{Cacciapaglia:2007jq, Galloway:2008jn} assumes that the Wilson line obeys
\be\label{Mand}
	\frac{\partial}{\partial y^\mu} W_M(z,y) = ie\,W_M(z,y)A_\mu(y) \;.
\ee
This is dubbed the `Mandelstam condition' following its application by that author to QED \cite{Mandelstam:1962mi}. A similar equation holds for $z^\mu$ from conjugation, and no path is specified for the line. Using (\ref{Mand}) one may expand the action (\ref{gauged-act}) in powers of the coupling and extract, for example, the one--gluon vertex \cite{Cacciapaglia:2007jq}
\be\label{terning-vert}
	\Gamma^a_\mu(p+q,p\,;q) = T^a\, \frac{2p_\mu +q_\mu}{2p.q+q^2}\big(\Delta^{-1}(p+q)-\Delta^{-1}(p)\big)\;,
\ee
where we define
\bestar
	(2\pi)^4\delta^4(p'-p-q)\Gamma^a_\mu(p',p\,;q) = \frac{1}{ie}\int\!\ud^4(z,y,x)\ e^{ip'\cdot z-ip\cdot y-iq\cdot x} \frac{\delta^3S}{\delta\phi^\dagger(z)\delta\phi(y)\delta A^\mu_a(x)}\bigg|_{A=0}\;.
\eestar
As a check, note that the vertex (\ref{terning-vert}) obeys the Ward--Takahashi identity
\be\label{ward}
	q^\mu\Gamma^a_\mu(p+q,p\,;q) = T^a\, \big(\Delta^{-1}(p+q)-\Delta^{-1}(p)\big)\;.
\ee
It was argued in \cite{Galloway:2008jn} that this method of gauging the unparticle action is the natural extension of the minimal coupling prescription to unparticle physics. To see this, note that since at $d=1$ the unparticle becomes a particle, we might expect that the vertices should then be those derived from the local, minimally coupled action
\bestar
	-\int\!\ud^4y\ \phi^\dagger(y)\, D^2\phi(y)\;,\qquad D_\mu \equiv \partial_\mu +ie A_\mu\;.
\eestar
This is the case for the vertex (\ref{terning-vert}), which reduces at $d=1$ to the usual three field vertex
\be\label{3}
	T^a\, (2p_\mu +q_\mu)\;.
\ee
It was shown in \cite{Galloway:2008jn} that the vertices of this theory also obey `generalised minimal coupling'  -- that is, when the scaling dimension is an integer, with $2-d\equiv n\geq 1$, the vertices are generated by the interaction $\phi^\dagger(D^2)^n\phi$.  We now give a new and rather simple proof of this: the inverse propagator at $2-d=n$ is
\bestar
	i\triangle(z-y)\bigg|_{2-d=n} =\frac{1}{a_n}\delta^4(z-y)\ (i\epsilon -\partial^2_y)^n\;.
\eestar
As an \emph{operator} expression, it follows from the Mandelstam condition (\ref{Mand}) that
\be\label{alld}
	\frac{\partial}{\partial y^\mu}\, W_M(z,y) = W_M(z,y)D_\mu(y)\;,
\ee
and similarly for \emph{any} number of derivatives. Using $W_M(y,y)=1$ the action (\ref{gauged-act}) then becomes
\be\label{nact}
	\frac{1}{a_n}\int\!\ud^4 y\ \phi^\dagger(y)(i\epsilon-D^2)^n \phi(y)\;,
\ee
at integer scaling dimensions, and thus generates the desired vertices.

Various aspects of the theory based on $W_M$ have been investigated -- for example the beta function \cite{Liao:2007fv, Basu:2008rd}, anomalies \cite{Galloway:2008sq}, unitarity and Ward identities \cite{Liao:2008dd}. Interestingly, the latter are found to be violated for on--shell (scattering) amplitudes. We will here focus on a more immediate problem -- the Wilson line with Mandelstam condition does not exist. As shown in \cite{LewisLicht:2008mq}, the best one could do is construct something similar with support only on pure gauge $A_\mu$. This follows from (\ref{Mand}) as we must have, multiplying by $W_M^{-1}$,
\be\label{Mand2}
	A_\mu(y)= \frac{1}{ie}W_M^{-1}(z,y)\, \frac{\partial}{\partial y^\mu} W_M(z,y)\;,
\ee
which is the condition that $A_\mu$ be pure gauge. (As a function of pure gauge fields, for which $F_{\mu\nu}$ vanishes, $W_M$ has found application to integrable models through the `zero curvature method', see Chap.~13 of \cite{Das:1989fn}.)

It seems that despite the mostly sensible results, in particular (\ref{terning-vert})--(\ref{3}), the above approach cannot really generate interactions between unparticles and arbitrary gauge fields $A_\mu$ -- in particular, it cannot couple to physical gauge fields, a point to which we will return in a later section.  Two solutions to this problem have been offered. The first is to go back to $S_\mathcal{U}$ and write $\Delta^{-1}$ as a sum over integer powers of $\partial^2$ in a derivative expansion \cite{LewisLicht:2008qr, Licht:2008km}. This essentially allows us to minimally couple by replacing (integer powers of) $\partial^2$ with $D^2$, arriving at an action which is an infinite sum over terms like (\ref{nact}),
\bestar
	S_\mathcal{U}\to \sum\limits_{n=0} E_n \int\!\ud^4y\ \phi^\dagger(y)(i\epsilon-D^2)^{n}\phi(y)\;.
\eestar
The definition of the coefficients $E_n$ is subtle and can be found in \cite{Licht:2008km}. To extract the Feynman rules one must expand each term of the action in the coupling and then resum. This requires some non--trivial complex analysis and differential--integral operator techniques, but reproduces the vertices of \cite{Cacciapaglia:2007jq, Galloway:2008jn} without any explicit use of $W_M$. It is possible that $W_M$ is simply a convenient shorthand for computing with these more technical operator methods, or that the problems with $W_M$ reappear as a subtlety of this approach. This is certainly worth further study.

In either case, there is a second, and very natural, proposed solution to the problem at hand -- this is to abandon $W_M$ and use instead a Wilson line with a specified path \cite{Licht:2008ic}. We turn to this approach below.

\subsection{The straight Wilson line action}
The only path for the Wilson line which preserves both the Poincar\'e and scale invariance of the unparticle action is a straight line between the unparticles \cite{LewisLicht:2008mv} (intuitively this makes sense as any other path must introduce a preference for some new direction and is then `less covariant' than the straight line). The resulting three field vertex, $\widetilde\Gamma$, may be compactly written \cite{Licht:2008ic}
\be\label{licht-vert-neat}
	{\widetilde\Gamma}^a_\mu(p\,;q):= -T^a\int\limits_0^1\!\ud s\ \frac{\partial}{\partial k^\mu}\Delta^{-1}(k)\bigg|_{k=-p-sq}\;,
\ee
although the explicit expressions for this and other vertices are more complex than those of \cite{Cacciapaglia:2007jq, Galloway:2008jn}. For example, we find (suspending the $i\epsilon$ prescription for a moment),
\be\label{licht-vert-nneat}
	{\widetilde\Gamma}^a_\mu(p\,;q) = T^a\frac{q_\mu}{q^2}\bigg[ ((p+q)^2)^{2-d} - (p^2)^{2-d}\bigg] + \,T^a\,(4-2d)\mathcal{A}^{1-d}C_{1-d}(p,q)\,t_{\mu\nu}(q)\,p^\nu\;,
\ee
where $t_{\mu\nu}(q) \equiv \delta_{\mu\nu} - q_\mu q_\nu/q^2$ is the momentum space transverse projector, $\mathcal{A}\equiv p^\mu t_{\mu\nu}(q) p^\nu$ and $C_{1-d}(p,q)$ is a hypergeometric function given in equations (27) and (29) of \cite{Licht:2008ic}. The Ward--Takahashi (\ref{ward}) identity is still obeyed, but now minimal coupling is recovered only when $d=1$, not for other integers \cite{Galloway:2008jn}. Nevertheless, the approach is unambiguous and well defined. It would be interesting to study the phenomenology of this theory, in particular the dependence on the specified path.

We will return to the roles of the Wilson line and the Mandelstam condition below, after we have understood more of the principles behind unparticle actions. This is the subject of the next section. As we will see, there are (infinitely) many objects which transform like the Wilson line, each of which will lead to a different gauged unparticle theory. For clarity we will initially restrict ourselves to U(1) gauge fields where everything can be done explicitly and exactly, returning to SU($N$) in Sec.~\ref{na-sec}.

\section{The most general U(1) coupling}\label{build-sec}
At the most basic level, gauging the unparticle action amounts to identifying a function $W(z,y)$ which transforms like a Wilson line, so that the action is gauge invariant. We will see that this requirement specifies only a minimal part of the unparticle action, and that all the physics is contained in the choices one makes to supplement this.

There is a large body of work addressing a closely related problem in the standard model, namely that of constructing gauge invariant charged and uncharged states. We will not go into details here, instead directing the reader to \cite{Lavelle:1995ty, Bagan1, Bagan2} for pedagogic and comprehensive reviews, but we will use the ideas of that approach to construct new actions and, later, discuss non--perturbative unparticle physics.
\subsection{The minimal action and the Mandelstam condition}\label{mansub}
Our first example will reveal the fundamental role of the Mandelstam condition discussed in Sec.~\ref{rev-sec}. Consider a general abelian $W(z,y)$, which we expand as a power series in $A_\mu$,
\bestar
	W(z,y) = \exp\bigg[ -ie\sum\limits_{n=1}^\infty \int \omega_{\mu_1\ldots\mu_n}(z,y; q_1,\ldots ,q_n)\ A^{\mu_1}(q_1)\cdots A^{\mu_n}(q_n)\bigg]\;.
\eestar
Gauge invariance of the action requires that $W(z,y)$ transforms as in (\ref{cod}). This implies each of the coefficient functions $\omega$ is transverse to the gauge field, in each index, except for $\omega_\mu$, the longitudinal part of which, $$\omega_\mu^L(z,y;q)\equiv \frac{q_\mu q_\nu}{q^2}\,\omega^\nu(z,y;q)\;,$$ is determined \emph{uniquely} by
\bestar
	q^\mu \omega_\mu = -i\big[ \e^{iq.z}-\e^{iq.y}\big] \implies \omega^L_\mu = \bigg[\frac{\e^{iq.z}-\e^{iq.y}}{iq^2}\bigg]\,q_\mu\;.
\eestar
Transforming to co-ordinate space, we can write this part of $W$ as
\be\label{minW}
	W_L(z,y)\equiv \exp\bigg[-ie \frac{\partial.A(z)}{\partial^2}\bigg]\ \exp\bigg[\,ie \frac{\partial.A(y)}{\partial^2}\bigg]\;.
\ee
Although there remains a great deal of freedom in the transverse terms of the action, gauge invariance fixes the longitudinal terms exactly. This $W_L$ must appear in all gauged unparticle actions; it is the \emph{only} contribution we are forced to include by gauge invariance. We therefore refer to the action
\bestar
	\int\!\ud^4z\ud^4y\ \phi^\dagger(z)\,i\Delta^{-1}(z-y)\,W_L(z,y)\,\phi(y)\;,
\eestar
made gauge invariant using only $W_L$, as the `minimal action'. One may check that $W_L$ preserves the Poincar\'e and scale invariance of the original unparticle action $S_\mathcal{U}$, being separately invariant under the pairs of transformations
\be\label{trans}\begin{split}
	A_\mu(x) &\to A_\mu(x+b)\;,\qquad z^\mu\to z^\mu-b^\mu,\quad y^\mu\to y^\mu-b^\mu\;,\\
	A_\mu(x) &\to \lambda A_\mu(\lambda x)\;,\qquad z^\mu\to z^\mu/\lambda\;,\quad y^\mu\to y^\mu/\lambda\;.
\end{split}\ee
The three field vertex, $\Gamma^L_\mu$, of the minimal action is particularly simple,
\be\label{min-vert}
	\Gamma^L_\mu(p+q,p\,;q) = \frac{q_\mu}{q^2}\big(\Delta^{-1}(p+q)-\Delta^{-1}(p)\big)\;,
\ee
and obeys the Ward--Takahashi identity (\ref{ward}) by construction. Minimal coupling is not recovered for integer scaling dimensions -- the reason is that $W_L$ depends only on the longitudinal part of the gauge field $A^L$,
\be\label{long-def}
	A_\mu^L(x) \equiv \partial_\mu \frac{\partial.A(x)}{\partial^2}\quad\iff\quad A_\mu^L(q) \equiv q_\mu \frac{q.A(q)}{q^2}\;.
\ee
and so cannot be used to describe the interaction of observable photons with unparticles; in particular we can work in a gauge (Landau) in which the interaction vanishes. Nevertheless, in any explicitly gauge invariant approach we see that $W_L$ is an essential part of unparticle--gauge field coupling\footnote{One may also see this from the Ward--Takahashi identity which implies that the longitudinal part of all admissible three field vertices is precisely $\Gamma^L_\mu$, as in (\ref{min-vert}). Indeed, one may verify that (\ref{min-vert}) is the $q$--longitudinal part of the vertex (\ref{licht-vert-nneat}) and of (\ref{terning-vert}) after decomposing into longitudinal and transverse pieces.}.

We now show how the minimal action is related to the Mandelstam condition (\ref{Mand}). Recall that the Wilson line with Mandelstam condition can only depend on \emph{pure gauge} fields. We will refer to this object as the `Mandelstam function'. In U(1), the pure gauge component of $A_\mu$ is equal to its longitudinal component $A_\mu^L$.
It is easy to check that our $W_L$ obeys
\be
	\frac{\partial}{\partial y^\mu}\, W_{L}(z,y) = ie\, W_L(z,y)\, A_\mu^L(y)\;,
\ee
and similarly for $z^\mu$, and $W_L$ is therefore exactly the U(1) Mandelstam function. We can recover its Wilson line description by noting that the leading derivative in (\ref{long-def}) allows us to write
\bestar
	W_L(z,y) = \exp\bigg[ -ie \int\limits_\gamma\!\ud w^\mu A_\mu^L(w)\bigg]\;,
\eestar
for an \emph{arbitrary} path $\gamma$ between $y$ and $z$. So, although the definition (\ref{Mand}) of $W_M$ originally proposed cannot be satisfied, we have found that, properly interpreted, the Mandelstam condition plays a fundamental role in unparticle--gauge field coupling; it defines the minimal and unique part of the action required by gauge invariance.
\subsection{The most general action}
From the above, it follows that terms other than $W_L$ used to gauge the unparticle action must be transverse. Hence the most general interaction is given by
\be
	W_\mathcal{F}(z,y) \equiv W_L(z,y)\ \exp\bigg( -ie\,\mathcal{F}[A^t;z,y]\bigg)\;,
\ee
for any real function $\mathcal{F}$ depending on the transverse field $A^t_\mu\equiv t_{\mu\nu}A^\nu = A_\mu - A_\mu^L$ (and antisymmetric under interchange of $z$ and $y$).  As $\mathcal{F}$ depends on gauge invariant fields, there is no gauge in which the unparticles and gauge fields decouple. The vertices in $W_\mathcal{F}$ are therefore non--trivial; it is $\mathcal{F}$ which really provides the physical content of the theory. For example, as $W_L$ maintains both Poincar\'e and scale invariance under (\ref{trans}), these properties extend to the theory generated by $W_\mathcal{F}$ only if they are respected by $\mathcal{F}$ itself.

The possible structures of $\mathcal{F}$ will be clearer if we restrict ourselves to functions \textit{linear} in $A^t_\mu$. The only vector we can contract this field with, which maintains translation invariance, is $z^\mu-y^\mu$. This leaves us a scalar degree of freedom which we call $f$. The most general linear $\mathcal{F}$ is then
\be\label{vv-gen}
	\mathcal{F} = -ie\int\!\frac{\ud^4q}{(2\pi)^4}\bigg[f(q\,;z-y)\e^{iq.z}+f^*(-q\,;y-z)\e^{iq.y}\bigg]\ (z-y)^\mu t_{\mu\nu}(q) A^\nu(q)\;.
\ee
To impose other physical properties we place constraints on $f(q\,;z-y)$, such as
\begin{align}\label{bbb}
\nonumber \displaystyle f(\lambda q\,;\lambda^{-1}(z-y))= f(q\,;z-y)\ \bigg\} &\text{ scale invariance,}\\
\left.\begin{array}{c}
	\displaystyle\lim_{z\to y}\ f(q;z-y)\e^{iq.z}+f^*(-q;z-y)\e^{iq.y} = \e^{iq.y} \\ \vspace{15pt}
	\displaystyle\lim_{z\to y}\ \frac{\partial}{\partial y^\mu}\ f(q;z-y)\e^{iq.z}+f^*(-q;z-y)\e^{iq.y} \propto q_\mu
\end{array}\right\} &\text{ minimal coupling at $d=1$.}
\end{align}
The minimal coupling conditions follow from imposing the operator identity
\bestar
	\lim_{z\to y}\partial^2_y\ W_\mathcal{F}(z,y) = D^2(y)\;,
\eestar
which is comparable to, but less restrictive than, the corresponding Mandelstam condition (\ref{alld}). We now give explicit examples of unparticle actions and their three field vertices.
\subsection{Example -- generalised Wilson lines}
Let us make the choice $f(q\,;z-y)=\kappa/iq\cdot(z-y)$ for $\kappa\equiv \kappa(q\cdot(z-y))$. We arrive at
\be\label{v-gen}
	W_\mathcal{F} = W_L\exp\bigg[-ie\int\!\frac{\ud^4q}{(2\pi)^4}\ \kappa\big(q\cdot(z-y)\big)\bigg[\frac{\e^{iq.z}-\e^{iq.y}}{iq.(z-y)}\bigg]\ (z-y)^\mu A_\mu^t(q)\bigg]\;.
\ee
This parameterisation describes a generalisation of the Wilson line, as we now show. Taking first $\kappa\equiv 1$, we recover the (Fourier transform of) the straight Wilson line, i.e.
\bestar
	W_\mathcal{F}\to  \exp\bigg[\ -ie\int\limits_C\!\ud w^\mu A_\mu(w)\bigg]\;,\qquad C:s\in[0,1]\to y^\mu+s(z^\mu-y^\mu)\;.
\eestar
When $\kappa$ has a nontrivial functional dependence, $W_\mathcal{F}$ contains a Wilson line which is smeared, or deformed, by the function $\kappa$, and additional longitudinal fields are introduced, proportional to $1-\kappa$, in order to maintain gauge invariance. Minimal coupling at $d=1$ is maintained for $\kappa(0)=1$. An example is given by $\kappa=\cos q\cdot(z-y)$, for which the one photon vertex becomes a combination of $\Gamma_\mu^L$ and $\widetilde\Gamma_\mu$,
\be\begin{split}
	\frac{1}{2}\widetilde\Gamma_\mu(p-q;q)&+\frac{1}{2}\widetilde\Gamma_\mu(p+q;q)\\
	&+\frac{q_\mu}{2\,q^2}\big[3\Delta^{-1}(p+q)-3\Delta^{-1}(p)-\Delta^{-1}(p+2q)+\Delta^{-1}(p-q)\big]\;.
\end{split}\ee
Seen from this perspective it is clear that the straight Wilson line is just one of a wide class of possible functions preserving the symmetries of the original unparticle action and which give minimal coupling at $d=1$. It is also clear that while gauge invariance allows us to understand the structure of the action, it is certainly not enough to identify it uniquely. Ideally one would like to invoke physical principles which further constrain its form, as in (\ref{bbb}).  The idea has a natural analogue in particle theory. When constructing states of matter and gauge fields, Gauss' law constrains the states to be gauge invariant, but this does not uniquely identify a state \emph{nor} make it physical. Other conditions, such as minimisation of the energy, must be added to specify the dynamics and produce a physically relevant state.

\subsection{Example -- including heavy charge velocities}\label{heavysub}
Adding other fields to the theory allows more parameters to enter the unparticle--gauge interaction. If, for example, we have integrated out a heavy particle moving with velocity $v_\mu$, the interaction can contain velocity dependent terms. The simplest non--trivial example is to add a term depending on $t_{\mu\nu}v^\nu$, i.e. we gauge the unparticle action using
\be
	W_L(z,y)\ \exp\bigg( -ie\int\!\frac{\ud^4q}{(2\pi)^4}\ (\e^{iq.z}-\e^{iq.y})\,i\,v^\mu  A^t_\mu(q)\bigg)\;,
\ee
which has the correct gauge transformation and which yields the simple three field vertex
\be
	\big(\Delta^{-1}(p+q)-\Delta^{-1}(p)\big)\bigg[\frac{q_\mu}{q^2} + t_{\mu\nu}(q)v^\nu\bigg]\;.
\ee
This interaction cannot be gauged away. We do not recover minimal coupling at $d=1$, but since we have integrated out other fields we should not expect to. The Ward--Takahashi identity is still obeyed and the action remains scale invariant.

\section{Coupling to SU($N$) gauge fields}\label{na-sec}
The preceding results are immediately generalisable to SU($N$) gauge fields in perturbation theory. The expressions we have given in Sec.~\ref{mansub} to Sec.~\ref{heavysub} are the lowest order terms in a coupling expansion of the corresponding SU($N$) formulae. In particular, the one--photon vertices we have derived above automatically become one--gluon vertices when multiplied by the generator $T^a$. To illustrate an efficient method of extending these results, we will construct the SU($N$) Mandelstam function to second order in the coupling. We will then go on to discuss some non--perturbative aspects of these constructions.
\subsection{Perturbative construction of the Mandelstam function}
As for the abelian case, the Mandelstam function may be written as a product\footnote{At least in perturbation theory -- see the following section.} of two terms depending on the pure gauge piece, $A_\mu^\text{PG}$, of a given gauge field $A_\mu$,
\bestar
	W_M(z,y) = M^{-1}(z)M(y)\;,
\eestar
with $M(y)$ obeying
\be\label{bob}
	\frac{\partial}{\partial y^\mu} M(y) = ie\, M(y)A^\text{PG}_\mu(y)\;.
\ee
Multiplying (\ref{bob}) by $M^{-1}$, it follows that $M$ is just the gauge transformation which maps the classical vacuum into $A_\mu^\text{PG}$. The pure gauge field may be constructed using the results of \cite{Lavelle:1995ty}, where an expression for the gauge invariant gluon field is given in perturbation theory. This allows us to write down the gauge non--invariant field and extract from it the pure gauge piece. To first order in the coupling the result is
\be
	A_\mu^\text{PG} = \partial_\mu \alpha_1 + e\bigg(  \partial_\mu\alpha_2 + \frac{i}{2}[\partial_\mu \alpha_1, \alpha_1]\bigg)+\ldots
\ee
where
\be
	\alpha_1 \equiv \frac{\partial.A}{\partial^2}\;,\qquad \alpha_2 \equiv i\frac{\partial_\sigma}{\partial^2}\bigg( [v_1,A_\sigma] + \frac{1}{2}[\partial_\sigma v_1,v_1]\bigg)\;.
\ee
We then find
\be
	M(y) = \exp\bigg(\ ie\, \alpha_1(y) + ie^2\,\alpha_2(y)+\ldots\bigg)\;.
\ee
To lowest order, this result agrees with the abelian result (\ref{minW}), as we would expect. We can also check to order $e^2$ that (\ref{bob}) holds and that under gauge transformations,
\be\label{dave}
	M(y)\to M(y)U(y)\;,
\ee
so that the resulting unparticle action is gauge invariant. Higher order terms for this and our previous results may be similarly constructed. The non--perturbative existence of $M$ and $W_M$ is the subject of the next section.

\subsection{Non--perturbative coloured unparticles}
We now address the construction of gauge invariant, coloured unparticles. Note that the unparticle field operator $\hat\phi$ can not, on its own, create physical unparticles from the vacuum, as it is not gauge invariant. In this paper we have identified objects $W$, depending on $A_\mu$, whose gauge transformation (like the Wilson line) makes the action invariant. It is clear that if we view these $W$ as functionals $W[\hat A]$ of the gauge field operator, they may be used to write down gauge invariant `unmeson' operators, $\mathcal{\hat O}_\text{mes}$, where
\be
	\mathcal{\hat O}_\text{mes} = \hat\phi^\dagger(z)\, W[\hat A\,](z,y)\,\hat\phi(y)\;.
\ee
Furthermore, if $W$ can be factorised into $M^{-1}(z)M(y)$, as in (\ref{bob}) and (\ref{dave}), then we may write down a colour charged, gauge invariant operator $\mathcal{\hat O}_\mathcal{U}$,
\be
	\mathcal{\hat O}_\mathcal{U} = M[\hat A\,](y)\,\hat\phi(y)\;,
\ee
which, it appears, creates a gauge invariant, coloured unparticle. (In the language of \cite{Lavelle:1995ty, Bagan1, Bagan2}, $M$ is a `dressing'.) One can picture the matter created by $\hat\phi$ surrounded by a cloud of glue created by $M$, but it is the composite, gauge invariant, object which is identified with an observable unparticle. Note also that, by construction, the scaling dimension of $\mathcal{\hat O}_\mathcal{U}$ is the same as that of $\hat\phi$.

It is useful to compare these statements with their equivalents for particles in the standard model. The argument above depends only on the gauge transformation properties of $\phi$, not its dynamics, and therefore also applies in particle theory where $\mathcal{\hat O}_\mathcal{U}$ would be a gauge invariant quark operator. Of course, this cannot be the whole story, as no coloured particles are observed in nature. The missing element comes from distinguishing between perturbative and non--perturbative effects in the theory. It is indeed possible to construct individual, gauge invariant quarks in perturbation theory, and this reflects the expected short range (weak coupling) physics of the strong force. However, non--perturbatively, operators like $\mathcal{\hat O}_\mathcal{U}$ \emph{cannot} be made gauge invariant -- if they could, there would exist a globally well defined gauge fixing in QCD \cite{Lavelle:1995ty, Ilderton:2007qy}, which it is known does not exist because of the Gribov ambiguity \cite{Gribov:1977wm, Singer:1978dk}.

The result is that, non--perturbatively, quarks are not physical and therefore not observable, in agreement with experiment, and if quarks cannot be observed, then they must be confined into hadrons -- although it is not possible to find the quark analogue of $\mathcal{\hat O}_\mathcal{U}$, the mesonic operator $\mathcal{\hat O}_\text{mes}$ may exist but \emph{without} being factorisable. We will see an explicit example of this below.

Similarly, by constructing objects like our $M$, we may write down perturbative expressions for gauge invariant unparticles in perturbation theory. However, going from particles to unparticles certainly does not remove Gribov copies from the Yang--Mills configuration space, so the same non--perturbative obstruction must exist which prevents us building a non--perturbative coloured unparticle.

This implies that confinement of colour charge also holds in unparticle physics; if unparticles are found to be a part of the standard model, only colourless `unhadrons' will be observable. This immediately poses a puzzle -- what do we mean by an unhadron in a theory which is supposedly exactly scale invariant? How can we have an unhadron scale? It may be that unhadrons are simply so different from hadrons that scale invariance is preserved for unparticle bound states, or that coupling to gauge fields will, ultimately, spoil the scale invariance of the unparticle theory. Progress in understanding these interesting issues will only come with further investigation of gauge invariant unparticle and unhadron states. It would also be very interesting to compare our construction with other approaches which treat the unparticle as a composite, see \cite{Sannino:2008nv}, and to examine how dressing fits in to those scenarios. These are topics for future research, but we will end this section with a few initial comments on a possible unmeson state.

\subsection{Wilson line phenomenology}
Consider the following operator, built from a path ordered straight Wilson line,
\be
	\hat\phi^\dagger(x)\ \mathbb{P}\,\exp\bigg[ -ie \int\limits_y^z \ud w^\mu \hat A_\mu(w)\bigg] \,\hat\phi(y)\;.
\ee
This provides an example of a colourless $\mathcal{\hat O}_\text{mes}$ which is not hampered by the Gribov ambiguity. It cannot be factorised into two pieces (or dressings) describing two individually gauge invariant unparticles -- instead it describes a composite unparticle object, an `unmeson'.

As well as the question of scales addressed above, there is also the question of whether this Wilson line operator creates a \emph{physical} state in the unparticle theory. This brings us back to the discussion of Sec.~\ref{build-sec}, where we observed that additional input is needed to select an $W$ with some particular property (be it for the action or a state) out of the myriad possible.

In the standard model one finds that mesonic states made gauge invariant using Wilson lines have unwanted properties. In U(1) such states are unphysical, being infinitely excited, and unstable\footnote{At least in the non--compact theory, see \cite{Heinzl:2007kx}.}. Similar results for SU($N$) states have been found in lattice simulations, where the overlap of the states with the ground state is seen to be very poor, and potentially zero in the continuum limit. All these properties follow directly from the localisation of the (chromo--) electromagnetic fields onto an infinitely thin line \cite{Heinzl:2007kx, Heinzl:2008tv}. Do these undesirable properties persist in unparticle physics, where the dynamics will be very different to those in QCD? Coming full circle, it is an open question how such properties manifest themselves when a Wilson line is used to gauge the unparticle action, and what phenomenological implications this could have.

\section{Conclusions}\label{con-sec}
We have seen that after satisfying the requirements of gauge invariance, an infinite number of physical degrees of freedom remain available to the gauged unparticle action. This implies a huge spectrum of possible physical theories in addition to the two generated by the approaches of \cite{Cacciapaglia:2007jq} and \cite{Licht:2008ic}.

Our methods allowed us to write down the most general interaction between unparticles and U(1) gauge fields. This generalises the straight Wilson line action \cite{Licht:2008ic} to a wide new class of actions. We also revealed the role of the Wilson line with Mandelstam condition. Properly defined, this is \emph{fundamental} to gauging the unparticle action as it defines the unique contribution required by gauge invariance. We were able to construct the Mandelstam function explicitly for U(1) gauge fields and in perturbation theory for SU($N$) fields. There is an opportunity to explore the links between this object and the application of pure gauge fields to integrable models, as described in \cite{Das:1989fn}.  Other natural extensions for future work are the analysis of unitarity, Ward identities, anomalies, etc, derived from the more general unparticle actions we have constructed, and further study of the operator approach used to circumvent the problems with the Mandelstam condition as originally proposed.

We have also described how the functions which make the non--local unparticle action gauge invariant may be promoted to operators creating gauge invariant, coloured unparticles in perturbation theory. We have seen that for SU($N$) fields such constructions are forbidden, \emph{non--perturbatively}, by global properties of the Yang--Mills configuration space, just as gauge invariant quarks are forbidden, non--perturbatively, in QCD. This leads to many questions concerning the form of the confinement mechanism in unparticle physics, and the nature of colourless unhadrons. In particular, how is a confinement scale reconciled with a scale invariant theory? We feel this is an extremely interesting topic for future study.

\subsection*{Acknowledgements}
Many thanks to Thomas Heinzl, Martin Lavelle and David McMullan for useful discussions. The author is supported by an IRCSET Postdoctoral Fellowship.


\begin{thebibliography}{99}

\bibitem{Georgi1}
  H.~Georgi,
  Phys.\ Rev.\ Lett.\  {\bf 98} (2007) 221601
  [arXiv:hep-ph/0703260].

\bibitem{Georgi2}
  H.~Georgi,
  Phys.\ Lett.\  B {\bf 650} (2007) 275
  [arXiv:0704.2457 [hep-ph]].

\bibitem{Cheung:2008xu}
  K.~Cheung, W.~Y.~Keung and T.~C.~Yuan,
  arXiv:0809.0995 [hep-ph].

\bibitem{Rajaraman:2008qt}
  A.~Rajaraman,
  arXiv:0809.5092 [hep-ph].


\bibitem{Cacciapaglia:2007jq}
  G.~Cacciapaglia, G.~Marandella and J.~Terning,
  JHEP {\bf 0801} (2008) 070
  [arXiv:0708.0005 [hep-ph]].

\bibitem{Licht:2008ic}
  A.~L.~Licht,
  arXiv:0801.0892 [hep-th].

\bibitem{Galloway:2008jn}
  J.~Galloway, D.~Martin and D.~Stancato,
  arXiv:0802.0313 [hep-th].

\bibitem{Mandelstam:1962mi}
  S.~Mandelstam,
  Annals Phys.\  {\bf 19} (1962) 1.

\bibitem{Liao:2007fv}
  Y.~Liao,
  arXiv:0708.3327 [hep-ph].

\bibitem{Basu:2008rd}
  R.~Basu, D.~Choudhury and H.~S.~Mani,
  arXiv:0803.4110 [hep-ph].

\bibitem{Galloway:2008sq}
  J.~Galloway, J.~McRaven and J.~Terning,
  arXiv:0805.0799 [hep-ph].

\bibitem{Liao:2008dd}
  Y.~Liao,
  arXiv:0804.4033 [hep-ph].


\bibitem{LewisLicht:2008mq}
  A.~L.~Licht,
  arXiv:0802.4310 [hep-th].

\bibitem{Das:1989fn}
  A.~K.~Das,
  World Sci.\ Lect.\ Notes Phys.\  {\bf 30} (1989) 1.

\bibitem{LewisLicht:2008qr}
  A.~L.~Licht,
  arXiv:0801.1148 [hep-th].



\bibitem{Licht:2008km}
  A.~L.~Licht and W.~Y.~Keung,
  arXiv:0806.3596 [hep-th].

\bibitem{LewisLicht:2008mv}
  A.~L.~Licht,
  arXiv:0805.3849 [hep-th].


\bibitem{Lavelle:1995ty}
  M.~Lavelle and D.~McMullan,
  Phys.\ Rept.\  {\bf 279} (1997) 1
  [arXiv:hep-ph/9509344].

\bibitem{Bagan1}
  E.~Bagan, M.~Lavelle and D.~McMullan,
  Annals Phys.\  {\bf 282} (2000) 471
  [arXiv:hep-ph/9909257].

\bibitem{Bagan2}
  E.~Bagan, M.~Lavelle and D.~McMullan,
  Annals Phys.\  {\bf 282} (2000) 503
  [arXiv:hep-ph/9909262].

\bibitem{Ilderton:2007qy}
  A.~Ilderton, M.~Lavelle and D.~McMullan,
  JHEP {\bf 0703} (2007) 044
  [arXiv:hep-th/0701168].


\bibitem{Gribov:1977wm}
  V.~N.~Gribov,
  Nucl.\ Phys.\  B {\bf 139} (1978) 1.

\bibitem{Singer:1978dk}
  I.~M.~Singer,
  Commun.\ Math.\ Phys.\  {\bf 60} (1978) 7.

\bibitem{Sannino:2008nv}
  F.~Sannino and R.~Zwicky,
  arXiv:0810.2686 [hep-ph].


\bibitem{Heinzl:2007kx}
  T.~Heinzl, A.~Ilderton, K.~Langfeld, M.~Lavelle, W.~Lutz and D.~McMullan,
  Phys.\ Rev.\  D {\bf 77} (2008) 054501
  [arXiv:0709.3486 [hep-lat]].

\bibitem{Heinzl:2008tv}
  T.~Heinzl, A.~Ilderton, K.~Langfeld, M.~Lavelle, W.~Lutz and D.~McMullan,
  Phys.\ Rev.\  D {\bf 78} (2008) 034504
  [arXiv:0806.1187 [hep-lat]].

\end{thebibliography}
\end{document}